
\magnification=1200
\hoffset=-.1in
\voffset=-.2in

\vsize=7.5in
\hsize=5.6in
\def\diag{\,{\rm diag}\,}
\tolerance 10000

\baselineskip 12pt plus 1pt minus 1pt
\pageno=0
\centerline{\bf GEOMETRIC GRAVITATIONAL FORCES}
\smallskip
\centerline{{\bf ON PARTICLES MOVING IN A LINE}\footnote{*}{This
work is supported in part by funds
provided by the U. S. Department of Energy (D.O.E.) under contract
\#DE-AC02-76ER03069, and by the Swiss National Science Foundation.}}
\vskip 24pt
\centerline{D. Cangemi and R. Jackiw}
\vskip 12pt
\centerline{\it Center for Theoretical Physics}
\centerline{\it Laboratory for Nuclear Science}
\centerline{\it and Department of Physics}
\centerline{\it Massachusetts Institute of Technology}
\centerline{\it Cambridge, Massachusetts\ \ 02139\ \ \ U.S.A.}
\vskip 1.5in
\centerline{Submitted to: {\it Physics Letters B\/}}
\vfill
\centerline{ Typeset in $\TeX$ by Roger L. Gilson}
\vskip -12pt
\noindent CTP\#2147\hfill October 1992
\eject
\baselineskip 24pt plus 2pt minus 2pt
\centerline{\bf ABSTRACT}
\medskip
In two-dimensional space-time, point particles can experience a geometric,
dimension-specific gravity force, which modifies the usual geodesic equation of
motion and provides a link between the cosmological constant and the vacuum
$\theta$-angle.
The description of such forces fits naturally into a gauge theory of
gravity based on the extended Poincar\'e group, {\it i.e.\/}
``string-inspired'' dilaton gravity.
\vfill
\eject
Consideration of $(1+1)$-dimensional --- lineal --- gravity, for pedagogical
and perhaps physical purposes, requires inventing gravitational dynamics; the
Einstein theory cannot be modeled owing to the vanishing of the two-dimensional
Einstein tensor.  Recent activity$^{1,\,2}$ in this area has focused on
scalar-tensor theories,$^3$ whose dynamical variables are the metric tensor
and an additional scalar field, while gravitational field equations involve the
Ricci scalar $R$ --- the single quantity encoding all local geometric
information in two dimensions.  Two models, governed by the following {\it
geometric\/} actions,
$$\eqalignno{I_1 &= \int d^2x \,\sqrt{-g}\,\eta (R-\lambda) &(1)\cr
I_2 &= \int d^2x\,\sqrt{-g}\, (\eta R-\lambda) &(2)\cr}$$
are especially noteworthy, in that both can be equivalently described by {\it
gauge theoretical\/} actions based on the de~Sitter group$^4$ for $I_1$, and on
the Poincar\'e group$^2$ with central extension$^5$ for $I_2$, the latter
describing ``string-inspired'' dilaton gravity.$^{1,\,2}$  Moreover, both
theories are obtained by various dimensional reductions from three
dimensions,$^{3,\,6,\,7,\,8,\,9}$
 and also $I_1$ and $I_2$ can be related to each
other by various singular limiting procedures.$^{5,\,7,\,9}$  [In (1) and (2)
$\eta$ is a scalar field and $\lambda$ the cosmological constant.]

Here we examine the interaction of a point particle with the gravitational
field.  The usual matter action for a particle of mass $m$ on the worldline
$x^\mu(\tau)$ is constructed from the arc length.
$$I_m = - m \int ds = - m \int d\tau \sqrt{ \dot x^\mu (\tau) g_{\mu\nu}
\bigl(x(\tau)\bigr) \dot x^\nu (\tau)}
\eqno(3)$$
[We set the velocity of light
 to unity and our Minkowski signature is $(+-)$; the overdot denotes
differentiation with respect to $\tau$, which parametrizes the worldline
in an arbitrary way.] However, the
$(1+1)$-dimensional setting provides another possibility, which is specific to
this dimension and which enjoys various interesting geometric
attributes.  Our purpose here is to examine this
point-particle --- gravity interaction, and in the second half of this Letter
to show how naturally it fits with the gauge theoretical formulation for $I_2$.

Let us begin by observing that the usual geodesic equation of motion,
modified by an additional force ${\cal F}$,
$${d\over d\tau}\ {1\over N} \dot x^\mu +{1\over N} \dot x^\alpha
\Gamma^\mu_{\alpha\beta} \dot x^\beta = {\cal F}^\mu\ \ ,\qquad N \equiv
{1\over m} \sqrt{\dot x^\alpha g_{\alpha\beta} \dot x^\beta}
\eqno(4)$$
will lose general covariance if ${\cal F}$ is {\it externally\/} prescribed.
($\Gamma$ is the Christoffel symbol evaluated on the worldline, as also is the
metric tensor.)
In order to maintain general
covariance ${\cal F}$ must be constructed from the dynamical variables of the
theory, and experience with electromagnetism (which is {\it not\/} included in
the above discussion) shows that general covariance is preserved when
${\cal F}_\mu$
has the tensorial properties of ${\cal F}_{\mu\nu}\dot x^\nu$, where ${\cal
F}_{\mu\nu}$
is an anti-symmetric, second-rank tensor.  In dimensions greater than two, such
a tensor cannot be constructed from particle and/or gravitational variables;
it arises when electromagnetic (or other gauge-field) degrees of freedom
are dynamically active.  However, in two dimensions gravitational variables
allow constructing the required tensor.
$${\cal F}_{\mu\nu} =-{\cal B} \sqrt{-g}\,\epsilon_{\mu\nu}\eqno(5)$$
Here $\epsilon_{\mu\nu}$ (also $\epsilon^{\mu\nu}$)
is the {\it numerical\/}, anti-symmetric tensor
density, $\epsilon_{01}=-1$ ($\epsilon^{01}=1$) and ${\cal B}$ is a constant
fixing the magnitude of the field.
Thus, the covariant, two-dimensional equation of
motion involving matter and metric variables and generalizing the usual
geodesic equation reads
$${d\over d\tau} \ {1\over N}\dot x^\mu + {1\over N}\dot x^\alpha
\Gamma^\mu_{\alpha\beta} \dot x^\beta + {\cal B} g^{\mu\alpha} \sqrt{-g}\,
\epsilon_{\alpha\beta} \dot x^\beta = 0 \eqno(6)$$

The additional interaction term in (6) is similar to that
arising from a constant external electromagnetic field in flat two-dimensional
Minkowski space-time.  Indeed, just as the latter preserves the Poincar\'e
symmetry of a non-interacting point particle in flat space-time,
so similarly our {\it covariantly\/} constant field
(5) respects general covariance. Moreover, as we now describe,
the force law  can be derived from a
novel contribution to the matter-gravity action.

Construction of the addition to the conventional matter-gravity action (3),
 which gives rise to our force, is geometrically subtle.
Consider the two-form ${1\over 2} {\cal F}_{\mu\nu} \,dx^\mu\,dx^\nu$,
proportional to the volume $V\equiv-{1\over 2}\sqrt{-g}\,\epsilon_{\mu\nu}
\, dx^\mu\, dx^\nu$, which may also be expressed in terms of the
{\it Zweibeine\/} $e^a_\mu$ as $-{1\over 2}\epsilon_{ab} e^a_\mu
e^b_\nu\,dx^\mu\, dx^\nu$.  Since $V$ is closed,
$dV = 0$, it is exact, at least locally.
$$V = da\eqno(7\hbox{a})$$
Equation~(7a) defines a one-form, whose components are also seen to satisfy
$${\partial a_\nu \over \partial x^\mu}(x) - {\partial a_\mu
\over \partial x^\nu}(x)
 = -\sqrt{-g(x)}\,\epsilon_{\mu\nu}\eqno(7\hbox{b})$$
Since the right-hand side of (7b) is a tensor, $a$ can be taken as a
vector.  The covariant action is now constructed from $a$.
$$I_{\cal B} = - {\cal B} \int d\tau\, a_\mu \bigl(x(\tau)\bigr)
\dot x^\mu(\tau)
\eqno(8)$$
Under coordinate redefinition, both $a$ and $x(\tau)$ change, and it
is straightforwardly verified that $I_{\cal B}$
is a scalar.
Note that the gauge ambiguity $a_\mu\to a_\mu + \partial_\mu\beta$, which is a
consequence of the defining equation (7b), changes the action only by
end-point contributions.

Variation of $I_{\cal B}$ with respect to $x(\tau)$
produces with the help of (7b)
the force in (6).
$${\cal F}_\mu = -{\delta I_{\cal B}\over \delta x^\mu(\tau)}
= -{\cal B}\sqrt{-g\bigl(x(\tau)\bigr)}\epsilon_{\mu\nu}
\, \dot x^\nu (\tau) \eqno(9)$$
Thus while no local expression for the action is available, its
$x(\tau)$-variation is
local, giving rise to an electromagnetic-like, velocity-dependent force, which
is free of gauge ambiguity and preserves general covariance.

Next, we examine the contribution from $I_{\cal B}$
to the energy-momentum tensor
$T$ which is a functional of the matter variable $x(\tau)$ and
function of the field argument $x$.
$$T^{\alpha\beta}\bigl(x(\tau)|x\bigr) = - {2\over \sqrt{-g(x)}}\
{\delta\over \delta g_{\alpha\beta}(x)} \left( I_m + I_{\cal B}\right)
\eqno(10\hbox{a})$$
Evaluating the metric variation of $I_{\cal B}$ is
problematic in the absence of an explicit
formula for $a$.  However, we can fix the additional term in the
energy-momentum tensor by requiring its covariant conservation.  In this way
$T$ is found to be
$$T^{\alpha\beta}\bigl(x(\tau)|x\bigr) = \int d\tau {1\over N(\tau)} \dot
x^\alpha(\tau)\dot x^\beta(\tau)
\delta^2(x-x(\tau))-{1\over 2}\Lambda\bigl(x(\tau)|x\bigr)
g^{\alpha\beta}(x)\eqno(10\hbox{b})$$
and we satisfy
$$D_\alpha T^\alpha{}_\beta = 0 \eqno(11)$$
provided that
$${1\over 2}\ {\partial \Lambda \over \partial x^\mu} \bigl(x(\tau)|x\bigr)
= -{\cal B}
\int d\tau \,\epsilon_{\mu\nu} \dot x^\nu (\tau) \delta^2(x-x(\tau))\eqno(12)$$
Equation (12) is easily solved in the parametrization where $x^0(\tau) =
\tau$, and one finds
$$\Lambda = -{\cal B}\varepsilon \left(x^1 - x^1(t)\right) + \lambda
\eqno(13)$$
where $\lambda$ is constant.  Thus the metric variation of $I_{\cal B}$ also
is local, giving rise to
a cosmological ``constant'' that jumps by $2{\cal B}$
as the particle's trajectory is crossed, a property that is independent of
the above parametrization choice

Although the matter dynamics can be presented without commitment to a specific
model for gravity, the new interaction (8) fits especially naturally with a
gauge theoretical formulation for $I_2$, which we now review.$^5$

The structure group for the gauge theory is the extended $(1+1)$-dimensional
Poincar\'e group with (anti-Hermitian) Lorentz generator $J$, and translation
generators $P_a$ that close upon commutation on a central element, $I$, which
commutes with $J$ and $P_a$.
$$\left[ P_a, J\right] = \epsilon_a{}^bP_b\ \ ,\qquad \left[ P_a, P_b\right] =
\epsilon_{ab} I \eqno(14)$$
[Latin letters refer to tangent space components and are moved by the flat
metric tensor $h_{ab} = \diag(1,-1)$.]  Gauge connection one-forms are
associated with the generators: {\it Zweibeine\/} $e^a$ with $P_a$,
spin-connection $\omega$ with $J$, and an additional connection $a$ with $I$,
giving
a Lie-algebra  valued connection one-form $A$,
$$A = e^a P_a + \omega J + aI \eqno(15)$$
and curvature two-form $F$.
$$\eqalign{F&= dA + A^2 = f^a P_a + fJ + gI\cr
&= \left( de^a + \epsilon^a{}_b \omega e^b\right) P_a + d\omega J + \left( da
+ {1\over 2} e^a \epsilon_{ab} e^b\right) I\cr}\eqno(16)$$
It is recognized that $f^a$ is proportional to the torsion
and $f$ to the curvature, while $g$ involves the Abelian field strength
associated with $a$, supplemented by the volume two-form.

A finite group transformation, generated by the gauge function $\Theta$,
$$\Theta = \theta^a P_a + \alpha J + \beta I \eqno(17)$$
produces the following transformations on the connections.
$$\eqalign{e^a & \longrightarrow \bar{e}^a = \left( {\cal M}^{-1}\right)^a_{\
b} \left( e^b + \epsilon^b{}_c \theta^c \omega + d\theta^b\right)\cr
\omega &\longrightarrow \bar{\omega}=\omega+d\alpha\cr
a &\longrightarrow \bar{a} = a - \theta^a \epsilon_{ab} e^b - {1\over 2}
\theta^2\omega + d\beta + {1\over 2} d\theta^a \epsilon_{ab}\theta^b
\cr}\eqno(18)$$
Here ${\cal M}$ is the finite Lorentz transformation with rapidity $\alpha$
$${\cal M}^a{}_b = \delta^a{}_b \cosh \alpha + \epsilon^a{}_b
\sinh\alpha\eqno(19)$$
The curvature components transform similarly to (18), except that the
inhomogeneous shifts involving gauge parameter differentials are absent.
Equivalently, one may collect the curvature components into a multiplet $F^A =
(f^a,f,g)$, which then transforms by the adjoint $4\times 4$ representation of
the extended Poincar\'e group.
$$\eqalignno{F^A \to \bar{F}^A &= \sum\limits^3_{B=0}
\left( U^{-1}\right)^A_{\ B} F^B &(20)\cr
U&= \left( \matrix{{\cal M}^a{}_b & -\epsilon^a{}_c\theta^c & 0
\cr\noalign{\vskip 0.2cm}
0 & 1 & 0\cr\noalign{\vskip 0.2cm}
\theta^c\epsilon_{cd} {\cal M}^d{}_b & -\theta^2/2 & 1\cr}\right) &(21)\cr}$$
The upper left $3\times 3$ block in $U$ comprises the adjoint representation
of the conventional Poincar\'e group with $-\epsilon^a{}_c \theta^c$ giving
the translation, while the fourth row and column arise from the extension.
In this representation, the extended algebra possesses a non-singular,
invariant inner product $h_{AB}$,
$$h_{AB} = \left( \matrix{ h_{ab} & \phantom{-}0 & 0\cr
0 & \phantom{-}0 & -1\cr
0 & -1 & \phantom{-}0\cr}\right)\eqno(22)$$
for which ${}^TUh\,U=h$; this allows raising and lowering indices
$(A,B)$.

A gauge invariant
gravity action, equivalent to $I_2$, is constructed with the help of
a Lagrange multiplier multiplet $\eta_A = \left( \eta_a, \eta_2,\eta_3\right)$,
taken to obey the coadjoint transformation law.
$$\eta_A\to \bar{\eta}_A = \sum\limits^3_{B=0} \eta_B U^B{}_A \eqno(23)$$
It is then easy to show that the gravitational dynamics of the gauge invariant
action $I'_2$
$$I'_2 = \int \sum^3_{A=0} \eta_AF^A\eqno(24)$$
and of $I_2$
coincide. Note in particular that no cosmological
constant is present in (24);  in the gauge theoretical formulation it is not a
parameter in the Lagrangian, but a value taken on by a variable, here
$\eta_3$.\footnote{$^\dagger$}{This mechanism
for generating the cosmological constant is here
dictated by the group structure.  In other contexts it has been inserted ``by
hand,'' see Ref.~[10].}

A {\it geometric\/} formulation of the gravity-matter
system is provided by $\left(1/2\pi G\right)I_2 +
I_m + I_{\cal B}$, where $G$ is the gravitational coupling strength.
 But we seek a {\it gauge theoretical\/} description; so a
gauge theoretical
point-particle action is now constructed, following a variation of the
Grignani and Nardelli method.$^8$  We shall arrive again at $I_m+I_{\cal B}$,
once a gauge freedom is fixed.

Let us begin with the particle action in flat space-time, where the vector
potential for the additional force is explicitly given by ${1\over 2}{\cal B}
\epsilon_{ab}x^b(\tau)$.
The matter action, in first-order form and in the absence of
curvature, reads
$$I_{\rm matter} = \int d\tau \left\{ \left( p_a - {1\over 2}
{\cal B}\epsilon_{ab}
x^b\right) \dot x^a + {1\over 2}N \left( p^2 - m^2\right)\right\} \eqno(25)$$
All quantities depend on the parametrization $\tau$; when $p$ and the
Lagrange multiplier $N$ are eliminated, $I_{\rm matter}$ coincides with
$I_m + I_{\cal B}$ in the absence of gravity.  The theory is Poincar\'e
invariant, but owing to the additional
interaction, the translation algebra acquires a central extension.

To couple $I_{\rm matter}$ to gravity, we view $p_a$ as the first two
components of a coadjoint tangent space
four-vector $p_A$, with vanishing last component
$p_3=0$; consequently $p_a$ transforms according to
$$p_a \to \bar{p}_a = p_b{\cal M}^b{}_a \eqno(26)$$
[analogous to $\eta_a$ in (23) with $\eta_3$ set to zero].  The position
variable $x^a(\tau)$ is replaced by the tangent space coordinate
$q^a(\tau)$, taken to be the first two components of an adjoint four-vector
$q^A$ with third component set to unity and fourth component to ${1\over 2}
\left(q^ah_{ab}q^b - c\right)$, where $c$ is a fixed number
(this implies $q^A h_{AB} q^B=c$).   The consequent transformation law reads
$$q^a \to \bar{q}^a = \left({\cal M}^{-1}\right)^a{}_b
\left( q^b + \epsilon^b{}_c
\theta^c\right)\eqno(27)$$
[analogous to $f^a$ in (20) with $f$ set to unity and $g$ to ${1\over 2}
\left(q^ah_{ab}q^b-c\right)$].  The time derivative
of the position variable is promoted to a covariant derivative
$$\left( D_\tau q\right)^a = \dot q^a + \epsilon^a{}_b \left( q^b\omega_\mu -
e^b_\mu\right)\dot x^\mu \eqno(28)$$
Here is introduced the worldline on the manifold $x(\tau)$; the
space-time indices in the gravitational potentials are
saturated with $\dot x^\mu$.  The
transformation law for the covariant derivative, which follows from (18) and
(27), is
$$\left( D_\tau q\right)^a \to \left({\cal M}^{-1}\right)^a_{\ b}
 \left( {\cal D}_\tau
q\right)^b \eqno(29)$$
Finally, further terms are added to achieve gauge invariance.
Implementing the above steps results in the following
action.\footnote{$^\ddagger$}{The various sign changes from (25)
are necessitated by the fact that the {\it
Zweibein\/} enters the symplectic term $pD_{\tau}q$ with a ``twist'' by
$\epsilon^a{}_b$; compare (41).}

$$I'_m + I'_B = \int d\tau\left\{ \left( p_a + {1\over 2} {\cal B}\epsilon_{ab}
q^b\right) \left( D_\tau q\right)^a - {1\over 2}N
\left( p^2 + m^2\right)-{\cal B}\left(
a_\mu - {1\over 2} q_a e^a_\mu \right) \dot x^\mu\right\} \eqno(30)$$
The dynamical quantities to be varied independently in (30) are the particle
variables
$p(\tau)$, $q(\tau)$ and $x(\tau)$,  the gravitational gauge
potentials $e$, $\omega$ and $a$, evaluated on $x(\tau)$,
and the Lagrange multiplier $N(\tau)$.

Performing the gauge transformations (18), (26), (27) and (29) on (30)
demonstrates that the action is invariant, apart from end-point contributions.
$$I'_m + I'_{\cal B}
\to I'_m + I'_{\cal B} - {\cal B}\int d\tau {d\over d\tau} \left( \beta -
{1\over 2} q_a \theta^a\right) \eqno(31)$$
We have not found a manifestly covariant formulation with non-vanishing
${\cal B}$.\footnote{$^{\dagger\dagger}$}{An alternative formula
for the matter Lagrangian, which uses
covariant notation as far as possible, is
$p_A\left(D_\tau q\right)^A - {1\over 2}
N \left( p_A h^{AB} p_B + m^2\right) +{1\over
2} {\cal B} \dot q^a\epsilon_{ab}q^b
+ {\cal B}\dot x^\mu A^A_\mu h_{AB} q^B$.}
Note that owing to the gauge variation (27) of $q^a$, one can always pass to a
gauge where $q^a$ vanishes, thereby simplifying the theory, see below.

The total action in the gauge group formulation
$$I = {1\over 2\pi G} I'_2 + I'_m + I'_{\cal B} \eqno(32)$$
is now varied to derive equations of motion.

Variation of the gravitational Lagrange multipliers $\eta_A$ determines the
geometry, and here matter variables do not enter.  We find, as in pure
gravity,
$$\eqalignno{0 &= {\delta I\over \delta\eta_a(x)} \Longrightarrow f^a = de^a +
\epsilon^a{}_b \omega e^b = 0 &(33)\cr
0 &= {\delta I\over \delta\eta_2(x)} \Longrightarrow f = d\omega = 0 &(34)\cr
0 &= {\delta I\over \delta\eta_3(x)} \Longrightarrow g = da + {1\over 2} e^a
\epsilon_{ab} e^b = 0 &(35)\cr}$$
The first of these enforces vanishing torsion and permits evaluation of the
spin-connection.
$$\omega = e^a \left( h_{ab} \epsilon^{\mu\nu} \partial_\mu
e^b_\nu\right)\big/\det e \eqno(36)$$
The second requires vanishing curvature, and is solved by
$$e^a_\mu = \delta^a_\mu\eqno(37)$$
which implies that $\omega$ in (36) vanishes.
Finally the third evaluates the vector potential $a$, precisely in the way
required by our particle gravity interaction, compare (7a).  With the trivial
geometry enforced by (33), (34), (36) and (37) we have (apart from a pure
gauge contribution)
$$\eqalign{a_\mu &= {1\over 2} \epsilon_{\mu\nu} x^\nu\cr
{\cal F}_{\mu\nu} &\equiv{\cal B}\left( \partial_\mu a_\nu - \partial_\nu a_\mu
\right) = - {\cal B}\epsilon_{\mu\nu}\cr}\eqno(38)$$

Next we examine the particle equations obtained by varying $x(\tau)$ and
$q(\tau)$ after eliminating $p(\tau)$ and $N(\tau)$.
We present these only at $q=0$, which results from a gauge choice, as
explained above.
$$\eqalignno{{\delta I\over\delta x^\mu(\tau)}\bigg|_{q=0}
&= 0 \Longrightarrow {d\over d\tau}\ {1\over N}\dot x^\mu + {1\over N} \dot
x^\alpha \Gamma^\mu_{\alpha\beta} \dot x^\beta = {\cal F}^\mu{}_\nu \dot x^\nu
&(39)\cr
{\delta I\over \delta q^a(\tau)}\bigg|_{q=0} &= 0
\Longrightarrow {1\over N} \dot x^\alpha \dot
x^\beta \bigl(  \partial_\alpha e^a_\beta - \Gamma^\mu_{\alpha\beta} e^a_\mu +
\epsilon^a{}_b \omega_\alpha e^b_\beta\bigr)
= -\dot x^\alpha \bigl( e^a_\mu {\cal F}^\mu{}_\alpha + {\cal B}\epsilon^a{}_b
e^b_\alpha\bigr) &(40)\cr}$$
We have used (39) to simplify (40), which is then
identically satisfied: the left
side vanishes (there is no torsion), the right side also vanishes when it is
recalled that ${\cal F}_{\mu\nu}
= - {\cal B} \sqrt{-g}\,\epsilon_{\mu\nu}$.  Thus, varying
$q$ does not produce an independent equation here, and $q$ may be set to
zero in the action, leaving a gauge fixed expression.
$$I\big|_{q=0} = {1\over 2\pi G} I'_2 + \int d\tau\,\biggl( p_a
\left( - \epsilon^a{}_b e^b_\mu \dot
x^\mu\right) - {1\over 2}N\left( p^2 + m^2\right) - {\cal B}a_\mu \dot x^\mu
\biggr)\eqno(41)$$
Since here $a$ satisfies (35), the matter part of (41) is the first-order form
of $I_m + I_{\cal B}$, given in (3) and (8).

Finally we record the equations for the gravitational Lagrange multipliers
$\eta_A$, which follow from varying the gravitational potentials, always at
$q=0$.
$$\eqalignno{{\delta I\over \delta e^a_\mu(x)}\bigg|_{q=0} &= 0 \Longrightarrow
\partial_\mu \eta_a + \epsilon_a{}^b \omega_\mu \eta_b + \eta_3 \epsilon_{ab}
e^b_\mu =2\pi G\int d\tau\,\delta^2(x-x(\tau)) {h_{ab} e^b_\alpha
\dot x^\alpha \epsilon_{\mu\nu}
\dot x^\nu\over N}\hskip .3in &\qquad(42)\cr
{\delta I\over \delta\omega_\mu(x)}\bigg|_{q=0}
&= 0 \Longrightarrow \partial_\mu\eta_2 +
\eta_a \epsilon^a{}_b e^b_\mu = 0 &(43)\cr
{\delta I\over\delta a_\mu(x)}\bigg|_{q=0}
&= 0 \Longrightarrow \partial_\mu\eta_3 =2\pi G \ {\cal B}
\int d\tau\,\epsilon_{\mu\nu} \dot x^\nu\delta^2(x-x(\tau)) &(44)\cr}$$
We recognize in (44) our previous equation (12), determining the
cosmological ``constant'' and solved in (13).
$${1\over 2\pi G}\eta_3 = {1\over 2} {\cal B}
\varepsilon\left(x^1-x^1(t)\right) -{1\over 2}\lambda \equiv - {1\over
2}\Lambda \eqno(45)$$

Owing to the trivial geometry (37), (38), the remaining
equations reduce as follows, in the parametrization $x^0(\tau)=\tau$.
The matter equation of motion (39) becomes
$${d\over dt} \  {m\dot x^\mu\over \sqrt{1-v^2}} = - {\cal B}
 \epsilon^\mu{}_\nu
\dot x^\nu\eqno(46)$$
[$v\equiv \dot x^1(t)$], and is solved by
$$x^1(t) = \bar{x}^1 + \sqrt{(t-\bar{x}^0)^2 + m^2/{\cal B}^2}  \eqno(47)$$
where $\bar{x}^\mu$ are integration constants
related to $P^\mu$, the energy $P^0$ and momentum, $P^1$.
$$P^\mu = {m\dot x^\mu(t)\over\sqrt{1-v^2}} + {\cal B}\epsilon^\mu{}_\nu x^\nu
(t) = {\cal B}\epsilon^\mu{}_\nu \bar{x}^\nu\eqno(48)$$
[Note that the energy is also given by the spatial integral of $T^{00}$ in
(10b), apart from an infinite quantity proportional to $\lambda$. However, the
spatial integral of $T^{01}$ is not $P^1$; $\int^\infty_{-\infty} dx^1 T^{01}$
is not time-independent even though $T^{\mu\nu}$ is conserved. Rather,
because $T^{11}$ remains non-vanishing at $x^1 = \pm\infty$, one must add
$t\int^\infty_{-\infty}dx^1 {d\over dx^1} T^{11}$ to $\int^\infty_{-\infty}
dx^1 T^{01}$ to achieve a time-independent quantity, which then coincides with
$P^1$.]

Equation (42) becomes, in view of  (45) and (47)
$$\partial_\mu \eta_a +\pi G\Lambda\epsilon_{\mu a}
=
{\cal T}_{\mu a}\eqno(49)$$
where ${\cal T}_{\mu a}$ is
$${\cal T}_{\mu a} = \pi G \left( P_a - {\cal B} \epsilon_{ab} x^b (t)\right)
\partial_\mu \varepsilon\left( x^1 - x^1(t)\right) \eqno(50)$$
This has the solution
$$\eta_a = \pi G \left( P_a - {\cal B}\epsilon_{ab} x^b\right)\varepsilon\left(
x^1-x^1(t)\right) + \pi G \lambda \epsilon_{ab} \left( x^b - x^b_0\right)
\eqno(51)$$
with $x^b_0$ arbitrary constants.

Finally, Eq.~(43) for $\eta_2$ reduces to
$$\partial_\mu \eta_2 + \eta_a\epsilon^a{}_\mu = 0 \eqno(52)$$
When $\eta_a$ is taken from (51), this is solved by
$$-2\eta_2 = M - \pi G\lambda\left( x^\mu - x^\mu_0\right)^2 - {\pi G\over
{\cal B}}\left( \left( P^\mu - {\cal B}\epsilon^\mu{}_\nu x^\nu\right)^2 -
m^2\right)
\varepsilon\left( x^1 - x^1(t)\right) \eqno(53)$$

In the ``string inspired'' gravity model,$^{1,\,2}$ the ``physical'' metric is
$g_{\mu\nu}/(-2\eta_2)$ and exhibits in the absence of
matter the geometry of a black hole with mass $M$ and location $x_0$.
We see that inclusion of matter, with the additional interaction that
we have introduced, merely shifts the black hole parameters.

We conclude with the following observations.
\medskip
\item{1)}The additional interaction results in an intriguing
connection between the gauge-theoretic $\theta$-angle characterizing the
vacuum state of the quantized theory and the cosmological
constant: In the gravity sector of the theory the $\eta_3-a$  interaction
gives rise to the cosmological ``constant'' proportional to $\eta_3$, on the
other
hand in the matter sector $a$ plays the role of a background electric
field, which is known to produce the $\theta$-angle in $(1+1)$-dimensional
gauge theories.$^{11}$
\medskip
\item{2)}The additional interaction gives a central extension to the
translation algebra in the matter sector, so that the Poincar\'e group is
realized in the same manner as in gravity. However, in the absence of this
interaction, when the Poincar\'e group action on the matter variables loses the
extension, the theory is till invariant; indeed, surface terms like those
in (31) are even absent.
Evidently invariance of a theory does not require that the
symmetry group be realized in the same way on all variables.
\medskip
\item{3)}In higher dimensions our construction is not available; an
alternative is the following:\footnote{$^{\ddagger\ddagger}$}{We
thank B.~Zwiebach for discussion on this point.}
 The closed
$d$-dimensional volume form defines
a $(d-1)$-form as in (7a).  The dual ${}^*a$ is then a one-form which can be
coupled to $\dot x^\mu$ in any dimension --- giving rise to force involving
$d{}^*a$.  However, such an interaction does not appear well-defined because
$a$ is determined by (7a) only up to an exact $(d-1)$-form (gauge ambiguity),
but such an ambiguity leaves $d{}^*a$ undetermined.
\vfill
\eject
\centerline{\bf REFERENCES}
\medskip
\item{1.}C. Callan, S. Giddings, A. Harvey and A. Strominger, {\it Phys. Rev.
D\/} {\bf 45}, 1005 (1992).
\medskip
\item{2.}H. Verlinde, in {\it Sixth Marcel Grossmann Meeting on General
Relativity\/}, M. Sato, ed. (World Scientific, Singapore, 1992).
\medskip
\item{3.}C. Teitelboim, {\it Phys. Lett.\/} {\bf 126B}, 41 (1983); in {\it
Quantum Theory of Gravity\/}, S. Christensen, ed. (Adam Hilger, Bristol,
1984); R. Jackiw, in {\it ibid.\/}; {\it Nucl. Phys.\/} {\bf B252}, 343
(1985).
\medskip
\item{4.}T. Fukiyama and K. Kamimura, {\it Phys. Lett.\/} {\bf 160B}, 259
(1985); K. Isler and C. Trugenberger, {\it Phys. Rev. Lett.\/} {\bf 63}, 834
(1989); A. Chamseddine and D. Wyler, {\it Phys. Lett.\/} {\bf B228}, 75
(1989).
\medskip
\item{5.}D. Cangemi and R. Jackiw, {\it Phys. Rev. Lett.\/} {\bf 69}, 233
(1992).
\medskip
\item{6.}D. Cangemi, MIT preprint CTP\#2124 (July 1992).
\medskip
\item{7.}A. Ach\'ucarro, Tufts University preprint (July 1992).
\medskip
\item{8.}G. Grignani and G. Nardelli, Perugia University preprint
DFUPG-57-1992 (August 1992).
\medskip
\item{9.}S.-K. Kim, K.-S. Soh and J.-H. Yee, Seoul University preprint
SNUTP-92-67 (1992).
\medskip
\item{10.}A.~Aurilia, H. Nicolai and P. Townsend, {\it Nucl. Phys.\/} {\bf
B176}, 509 (1980).
\medskip
\item{11.}S.~Coleman, {\it Ann. Phys.\/} (NY) {\bf 101}, 239 (1976).
\par
\vfill
\end